\documentclass[a4paper,amsfonts, amssymb, amsmath, reprint, showkeys,twoside,onecolumn,pra]{revtex4-1}
\usepackage[english]{babel}
\usepackage[utf8]{inputenc}
\usepackage{xcolor}
\usepackage{color}
\usepackage[colorinlistoftodos, color=green!40, prependcaption]{todonotes}
\usepackage[pdftex, pdftitle={Article}, pdfauthor={Author}]{hyperref} 

\begin{document}
\title{Perfect, Pretty Good and Optimized Quantum State Transfer in Transmon qubit chains}

\author{Pablo Serra$^{(1)}$}
\author{Alejandro Ferr\'on$^{(2)}$}
\author{Omar Osenda$^{(1)}$}
    \email[Correspondence email address: ]{osenda@famaf.unc.edu.ar}
    \affiliation{${(1)}$ Instituto de Física Enrique Gaviola, Universidad Nacional de Córdoba, CONICET, Facultad de Matemática, Astronomía, Física y Computación, Av. Medina Allende s/n, Ciudad Universitaria, CP:X5000HUA Córdoba, Argentina.\\ ${(2)}$ Instituto de Modelado e Innovación Tecnológica (CONICET-UNNE) and Facultad de Ciencias Exactas, Naturales y Agrimensura, Universidad Nacional del Nordeste, Avenida Libertad 5400, W3404AAS Corrientes, Argentina.}

\date{\today} 

\begin{abstract}
Chains of transmon qubits are considered promising systems to implement different quantum information tasks. In particular as channels that perform high-quality quantum state transfer. We study how changing the interaction strength between the chain qubits allows us to obtain perfect or pretty good state transfer and present explicit analytic expressions for their transmission fidelity. For particular values of the interactions between the qubits, transmon chains are equivalent to generalized SSH chains and show the traditional traits observed in chains with topological states, localized states at the extremes of the chain, and eigenvalues that lie inside the spectral gap. Consequently, we study the quantum state transfer on chains with dimerized interactions, looking for chains with fast transfer times. We show that, in many cases, asking for fast transfer times results in chains with dimerized interactions that do not have topological states.
\end{abstract}


\maketitle

\section{Introduction}
\label{sec:introduction}

The transmission of quantum states over chains of spins, or more generally, two-state systems, is one of the tasks of quantum state processing that have received continuous attention over the last twenty years \cite{Chang2023,Keele2022,Xie2023,Maleki201,Serra2022JPA,Mograby2021,Christandl2004}.  In the beginning, the area was academic, considering chains made of tens of components were clearly beyond the reach of the available technology.  

Two decades later, the scenario is different since quantum simulators and processors composed of tens of components are viable using diverse implementations and technologies. The most improved setups amongst those based on condensed matter physics are probably the ones made with superconducting qubits. After Nakamura and collaborators proposed the Cooper-pair box as a qubit \cite{Nakamura1999}, there have been numerous proposals of qubits based on superconductor materials, which resulted in a dramatic improvement in the decoherence time of almost six orders of magnitude \cite{Krantz2019}. One of the most interesting designs is the so-called transmon \cite{Koch2007,Schereier2008,Stehlik2021}.  

Processors or chains made of transmon qubits are achievable with present technology, making them the subject of many studies ranging from implementing quantum algorithms to effective quantum state transfer \cite{Kim2023,GoogleAI2023}. The couplings amongst the qubits are tunable site-by-site, in strength and time, so some transmission quantum state protocols involve the adiabatic change of the coupling's value to send a given quantum state from one point of the chain to another. The Hamiltonian of a dimerized transmon qubit chain looks like the Su-Schrieffer-Hu model one and presents topological states. The availability of these topological states suggests that coding the information in them would prevent the loss of information since they lie on the energy gap, which protects them from different mechanisms of decoherence and the presence of static disorder \cite{Estarellas2017,Lemonde2018,Lemonde2019,Mistakidis2020,Yousefjani2021a,Yousefjani2021b}.

Although the potential use of topological states in chains seems appealing, it has serious drawbacks. First, although the time control of individual interaction couplings is achievable, the simultaneous and precise control over a large set of interaction couplings is far from given. Second, the adiabatic change of the interaction values requires that the quantum state transfer proceeds over exceedingly long time intervals \cite{Estarellas2017,Wang2022,Serra2024}. The longer the arrival time of the transmitted state, the larger the chance that time-dependent disturbances affect the transmission. 

Because of the arguments in the paragraph above, it is compelling to look for time-independent Hamiltonians realizable in transmon qubit chains that result in high-fidelity quantum state transfer at finite and known arrival times. In a single-shot experiment, the arrival time is the elapsed time between the preparation of the state and the measurement of the transmitted state at the receiver's point of interest. Most transmission protocols try to maximize the fidelity of transmission value at a given known time. Ideally, the fidelity value should reach unity, although when perfect transmission is not possible, a compromise between large values of fidelity and short transmission times is advisable.

Recently, a particular architecture for transmon qubit chains has attracted some attention \cite{Wang2022}. In this architecture, the chain consists of a linear array of unit cells with three qubits each, but the last cell contains only two. The three qubits in each cell are not equivalents and are called the $A_1$, $A_2$, and $B$-type. The qubits type $A_2$ interact only with those of $A_1$ type, while the qubits type $A_1$ interacts with the qubit $B$ located in the same cell and with another qubit $B$ in an adjacent cell. Following this notation, it is customary to denote the strength of the interaction between the qubits $A_1$ and $A_2$ as $g_i$, the intra-cell interaction between the qubits $A_1$ and $B$ as $v_i$, and the interaction inter-cell between qubits $B$ and $A_1$ as $w_i$, where $i$ is the index numbering of the cells. In this context, the Hamiltonian with $v_i=v$, $w_i=w$, and $g_i=g$  corresponds to a generalized SSH model, with dimerized interactions $v$ and $w$. 

Chains like the one described above show good transfer ability for very long times, but their properties are known for limited sets of length and interaction values. In this paper, we show that considering different configurations for the interaction strength values, it is possible to construct chains that present pretty good quantum state transfer (PGT) \cite{Serra2024,Burgarth2006,Godsil2012a,Godsil2012b,Vinet2012,Kay2010,Banchi2017,Godall2012}, perfect state transfer (PT) at finite arrival times \cite{Christandl2004,Coutinho2016,6hristandl2005,Burgarth2005,Bayat2014}, and fast and efficient transmission with very high fidelity at short known arrival times \cite{Banchi2010,Zwick2015}. 

We organized the paper as follows. Section~\ref{sec:model} presents the model, its Hamiltonian, and the figure of merit quantifying the transfer process. In Section~\ref{sec:exact-and-general} we present some lemmas, which state properties of the spectra of different chains models that will be studied in the following Sections.  We dedicate Section \ref{sec:pretty-good} to the exact analytical solution of a family of chains allowing pretty good state transfer. Although the utility of the  PGT scenario is debatable, we analyse the scaling of the arrival time with the number of components of the chain. Section \ref{sec:perfect} is devoted to the exact solution of a family of chains showing perfect transmission. To achieve perfect state transfer, the strength of the couplings of these chains must be site dependent 
\cite{Loft2016,Banchi2011,Chapman2016,Kandel2019,Kandel2021,Martins2017,Baum2022,Kostak2007,Zajac2016,Li2018}. Sections \ref{sec:upper} and \ref{sec:optimized-couplings} are devoted to the transfer performance of different coupling setups. In Section \ref{sec:upper}, we pay attention to the effect of several schemes with dimerized interactions and compare their performance against chains with non-dimerized chains. We show that the results found in Section \ref{sec:pretty-good} for the maximum value of the transmission probability attainable for homogeneous chains are an upper bound for the transmission probability of dimerized chains. In Section \ref{sec:optimized-couplings}, we use optimization techniques to find coupling setups that produce chains that show excellent transmission ability and, in some cases, perfect transmission.   In Section \ref{sec:conclusions}, we discuss and summarize our results. 

\section{The chain model}
\label{sec:model}

\begin{figure}[h]
\includegraphics[width=0.7\linewidth]{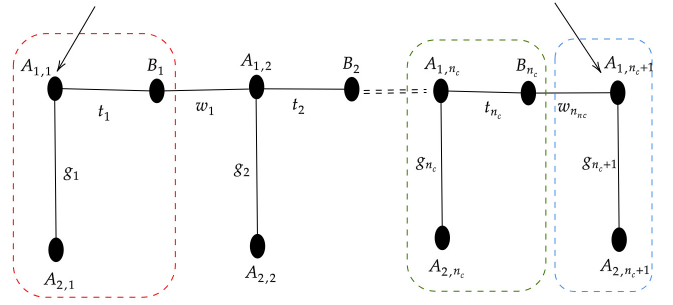}
\caption{The cartoon depicts all the ingredients of a decorated transmon chain in its most general form. The black circles represent the qubits on the system, and the solid black lines represent the first neighbour interactions between them. The three types of qubits, labelled as $A_{1}$, $A_{2}$, and $B$, reflect that an actual chain could consist of transmon qubits, couplers and controllers, which are two-level systems. The couplings between different kinds of qubits, $t_i$ between qubits type $A_{1,i}$ and $B_i$,  $w_i$ between qubits type $B_i$ and $A_{1,i+1}$, and $g_i$ between $A_{1,i}$ and $A_{2,i}$, are numbered accordingly with the cell-index $i$. In some situations, for instance, when considering dimerized systems, it is useful to group the qubits on the chain in cells with three qubits. The red dashed box to the left of the cartoon contains the three qubits that form the first cell. Note that all the labels corresponding to couplings or qubits have the same subindex, $1$. The green dashed box contains the last cell, and the blue dashed box contains only two qubits. As the number of cells is $n_c$, the total number of qubits is $3 n_c+2$. The two arrows in the top superior corners of the cartoon point to the "corner qubits". We analyse the quantum state transfer between the corner qubits.   }
\label{fig:ssh-decorada}
\end{figure}

We consider a chain made of superconducting qubits. In these chains, some qubits control the others, and there are implementations with three different kinds of qubits. The cartoon in Figure~\ref{fig:ssh-decorada} depicts a chain with these characteristics.

The  Hamiltonian for the decorated qubit chain \cite{Wang2022} is given by
\begin{equation} \label{ec:ssh-hamiltonian}
H= \sum_{i=1}^{n_c} \left( t_i \sigma_{A_{1, i}}^+ \sigma_{B_i}^- + w_i \sigma_{A_{1, i+1}}^ +\sigma_{B_i}^- + \mbox{h.c.} \right) + \sum_{i=1}^{n_c+1} \left( g_i \sigma_{A_{1, i}}^+ \sigma_{A_{2, i}}^-  + \mbox{h.c.} \right) \, ,
\end{equation}
where the strength of the nearest-neighbour interactions $g_i,w_i$, and $t_i$ can be tuned site by site.

The Hamiltonian in Eq.~\eqref{ec:ssh-hamiltonian} commutes with the total number of excitations operator, so it is natural to study the dynamics of quantum states restricted to spaces with a fixed number of excitations. Using this fact and that the state with no excitations is an eigenstate of the Hamiltonian, it is simple to show that the fidelity of the transmission of arbitrary one-qubit quantum states depends only on the transition probability 
\begin{equation}
P_{BA}(t) = |\left\langle \mathbf{1}_B \right| U(t) \left|  \mathbf{1}_A \right\rangle |^2 ,
\end{equation} 
where the one-excitation states $\left|  \mathbf{1}_A \right\rangle$, $\left|  \mathbf{1}_B \right\rangle$ correspond to a single excitation localized in the $A$ or $B$ sites of the chain, respectively.  The time-evolution operator, $U(t)$, is given by
\begin{equation}
U(t) = \exp{(-i h_N t)} ,
\end{equation}
where $h_N$ is the one-excitation restriction of the Hamiltonian in Eq.~\eqref{ec:ssh-hamiltonian} to the one-excitation sub-space, and $t$ is the time elapsed since the excitation dwelled in site $A$. Often, the performance of the protocol used to transfer arbitrary one-qubit quantum states is analysed using  the averaged transmission fidelity
\begin{equation}
f(t) = \frac{1}{2} + \frac{\sqrt{P(t)}}{3} \cos{\gamma}+\frac{P(t)}{6} ,
\end{equation}
using appropriate external control fields applied to the chain, the phase-dependent term $\cos{\gamma}=1$. 

There are many different combinations if the possible values of the coefficients are not somehow restricted. Some authors have considered dimerized interactions where $t_i=v$, $w_i=w$, and $g_i=g$, with $t\neq w$. This configuration is termed a decorated $SSH$ model and shows some physical traits found in the SSH model. 

 The model with all interactions equal, $t_i=w_i =g_i=1 $ has been thoroughly studied and shows good transfer abilities in asymptotic and pretty good state transfer PGT regimes, but the extremely long times involved in the transfer process preclude its use in actual implementations.  

\begin{figure}[h]
\includegraphics[width=0.7\linewidth]{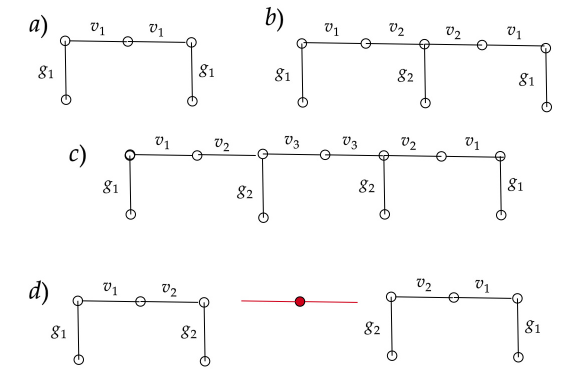}
\caption{The cartoon depicts three different symmetric chains. For symmetrical chains, there are two sets of couplings labelled $g_i$ and $v_i$, respectively. The symmetric chains have symmetric one-excitation Hamiltonians when written on the appropriate basis. From a), b) and c), it is clear that on a chain with $N=3k+5$ qubits, there are $k+2$ g-qubits that "hang" from a linear chain with $2k+3$ qubits. There are also $k+1$ different $v_i$ coupling coefficients. If $k$ is an even number, then a symmetric chain has $k/2+1$ different $g$ couplings, while if $k$ is an odd number, then the number of different g-couplings is $(k+1)/2$. d) Two symmetric chains glued together using a qubit, and the corresponding interactions generate a longer bisymmetric chain. Starting from two chains whose Hamiltonians have exact analytical algebraic expressions for all their eigenvalues, we use the glueing mechanism to produce chains whose Hamiltonian also has this property.    }
\label{fig:symmetric-chain}
\end{figure}

 When we look for chains with one-excitation Hamiltonian exactly solvable, we restrict our study to chains which are mirror symmetric concerning the centre of the chain, as the one depicted in the cartoon shown in Fig.~\ref{fig:symmetric-chain}. It is worth mentioning that we look not only for exactly solvable chain Hamiltonians but also want to obtain algebraic solvable expressions for the eigenvalues, which allow us to conduct a detailed study of the transfer properties of a given chain. 

Figure~\ref{fig:symmetric-chain}  a), b), and c) show the three non-trivial shortest symmetrical chains, with $N=5, 8$, and $11$ qubits. The total number of qubits in symmetric chains is  $N=3k+5$, where $k=0,1,2,3, \ldots$. For the case of the spin chains depicted in Figure~\ref{fig:symmetric-chain} a), b), and c) we have $k=0, 1$ and $2$, respectively. Figure~\ref{fig:symmetric-chain} d) shows how to obtain bi-symmetric chains by glueing two reflected chains using a single spin and two bonds. This mechanism allows us to construct larger chains from shorter ones. 

The exact form of the one excitation Hamiltonian matrix for a chain with $N$ spins depends on the interaction coefficients and the ordering of the qubits, see Appendix \ref{ap:numb}.

\section{Exact results on symmetric transmons chains}
\label{sec:exact-and-general}

In this section, we present some general results that hold for symmetric and bisymmetric qubit chains. 

\hspace{1ex}

\noindent {\bf Lemma 1}. All the eigenvalues of the Hamiltonian of a symmetric chain on $N$ qubits restricted to the subspace of one excitation, $h_N^s$, are also eigenvalues of $h_{2N+1}^s$.

\hspace{1ex}

\noindent {\bf Proof}. The demonstration proceeds using the Laplace theorem and writing the cofactor expansion of $det\left[h_{2N+1}^s -\lambda I\right]$ along the $N+1$ row.

\hspace{1ex}

\noindent {\bf Lemma 2}. Let it be $N=3k+5$, the length of a given chain, then zero is an eigenvalue of $h_N^s$ with multiplicity $k+1$.

\hspace{1ex}

\noindent {\bf Proof}. A chain of length $N=3k+5$ has $k+2$ qubits that interact with only another neighbour qubit. We call these qubits the "g-qubits". The chain also has $k+1$ qubits that do not interact with any g-qubit. Consequently, the chain has other $k+2$ qubits located at both sides of the qubits that do not interact with the g-qubits. It is easy to see that the file of the Hamiltonian matrix corresponding to a qubit that does not interact with a g-qubit is a linear combination of the files corresponding to the two qubits surrounding it. So, the rank of the Hamiltonian matrix is $Rank\left[ h_N^s \right]=N-(k+1) $ or, equivalently, the Hamiltonian matrix has $k+1$ null eigenvalues.

\hspace{1ex}

\noindent {\bf Lemma 3}. The nonnull eigenvalues of $h_N^s$ always appear in pairs of the for $\pm \lambda$. 

\hspace{1ex}

\noindent {\bf Proof}. The proof follows the proof of this result for $XY$ spin chains, see Reference \cite{Godsil2012a} and references therein.

\hspace{1ex}

As a consequence of lemmas 1, 2, and 3, $h_{2N+1}^s$ has only $(N+1)/3=k+2$ new eigenvalues concerning $h_{N}^s$.

\hspace{1ex}

\subsection*{Bisymmetric chains}.

\hspace{1ex}

Chains with an odd number of qubits are candidates to be bisymmetrical. We assume that we have a bisymmetric chain with $N$ qubits. Then, we construct another bisymmetric chain using the glueing mechanism, whose length is $2N+1$. The eigenvalues of the Hamiltonian $h_{2N+1}^b$, which corresponds to a bisymmetric chain with $2N+1$ qubits,  are closely related to the eigenvalues of $h_{N}^b$, as stated in the following lemma.

\hspace{1ex}

\noindent {\bf Lemma 4} All the eigenvalues of the Hamiltonian $h_N^b$ are also eigenvalues of the Hamiltonian $h_{2N+1}^b$. The remaining $N+1$ eigenvalues of $h_{2N+1}^b$ are eigenvalues of a matrix $N+1 \times N+1$, $A_{N+1}$, written in terms of $h_N^b$. The eigenvectors with zero eigenvalues of $A_{N+1}$ are related to the eigenvectors of $h_N^b$ with zero eigenvalues.

\hspace{1ex}

\begin{figure}[h]
\includegraphics[width=0.7\linewidth]{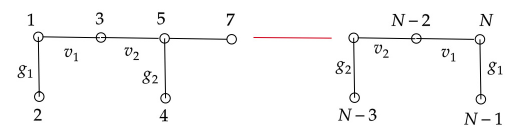}
\caption{The cartoon shows the numbering used for constructing the Hamiltonian matrix of a chain with $N$ qubits, which is required in the proof of the Lemma $4$ about the spectrum of bisymmetric chains. Note that the labels of the g-qubits are even numbers, while the other qubits have odd numbers as labels. The total number of qubits is odd.  }
\label{fig:numbering-bisymmetric}
\end{figure}

\noindent {\bf Proof}. To prove the lemma, we resort to a numbering for the qubits on a chain as shown in Figure \ref{fig:numbering-bisymmetric}.

Using this numbering, we can write the one excitation Hamiltonian as
\begin{equation}
h_{2N+1}^b=  \left(
\begin{array}{ccc}
h_N &  \vec{v}_l^{\dagger} & 0_N  \\
\vec{v}_l & 0 & \vec{v}_f   \\
0_N  & \vec{v}_f^{\dagger} & h_N
\end{array}
\right),
\end{equation}
where $0_N$ is an $N\times N$ matrix whose all elements are zero, $\vec{v_l}$ and $\vec{v}_f$ are two file vectors with $N$ components given by
\begin{equation}
\vec{v}_l = \left( 0,0,\ldots,0,v_{(N+1)/2}\right)
\end{equation}
and 
\begin{equation}
\vec{v}_f = \left(v_{(N+1)/2},0,0,\ldots,0\right) .
\end{equation}

Using well-known results from linear Algebra \cite{compu-mathema}, the one excitation Hamiltonian $h_{2N+1}$ can be block diagonalized as
\begin{equation}
D^{\dagger} h D = \left(
\begin{array}{cc}
A_{N+1} &   0  \\
0  &   h_N
\end{array}
\right),
\end{equation}
where the matrices $h_N$ and $A_{N+1}$ are symmetrical, but not bisymmetrical, except when $v_i=v, \forall i\neq (N-1)/3$. In this last case, $h_N$ is bi-symmetrical, and $N$ is an odd number. $A_{N+1}$ is given by
\begin{equation}
A_{N+1} = \left(
\begin{array}{cc}
h_N  &  \vec{v}_{N+1}^{\dagger}  \\
\vec{v}_{N+1}  &  0
\end{array}
\right),
\end{equation}
where 
\begin{equation}
\vec{v}_{N+1} = (0,\ldots,0,\sqrt{2} v_{(N-2)/3},0) ,
\end{equation}
and the transformation matrix is given by
\begin{equation}
D_{2N+1} = \frac{1}{\sqrt{2}} \left(
\begin{array}{ccc}
I_N  &  0  &  I_N  \\
0  &  \sqrt{2}  &  0  \\
S_N  &  0  & -S_N
\end{array}
\right),
\end{equation}
where the $I_N$ is the $N\times N$ identity matrix, and the $N\times N$ $S_N$ matrix is given by
\begin{equation}
S_N = \left(
\begin{array}{ccccccc}
0 & 0 & 0 & 0  &  \ldots  & 0 & 1 \\
0  &  0  &  0  & \ldots  &  0  & 1 & 0 \\
0  &  0  &  \ldots  & 0  &  1  & 0 & 0 \\
\vdots & \ddots  &\ddots  &\ddots  &\ddots  &\ddots  & \vdots \\
0  & 0  & 1  & 0 &\ldots &  0  &  0    \\
0  & 1  & 0  & \ldots  &  0  &  0  &  0  \\
1  &  0  & \ldots  & 0  &  0  &  0  &  0  
\end{array}
\right).
\end{equation}

Note that $h_N \vec{x}_N =0$ implies that the $N-$th component of the vector is null, so $x_N=0$. This last equation holds for all the eigenvectors of $h_N$ with null eigenvalue. 

Then, using the explicit expression of $A_{N+1}$, it is simple to show that if $\vec{x}_N= (x_1, \ldots,x_N)$ is an eigenvector with null eigenvalue of $h_N$, then the vector $(x_1, \ldots,x_N,0)$ is an eigenvector with null eigenvalue of $A_{N+1}$.

\section{Transmon qubit chains with pretty good state transfer}
\label{sec:pretty-good}

The pretty good quantum state transfer scenario described in numerous papers implies that by waiting long enough times, the fidelity of the transferred state would eventually become very close to the unity. Of course, this is not the best scenario for actual quantum state transfer since very long waiting times imply that different decoherence mechanisms have more opportunities to spoil the transmission. The same argument applies to asymptotic transmission regimes. 

Notwithstanding the arguments above, in this Section, we look for chains with exact solutions that show a pretty good state transfer scenario. In a previous paper\cite{Serra2022JPA}, we showed that the one-excitation Hamiltonians of short homogeneous chains are exactly solvable and obtained explicit expressions for the eigenvalues and the transfer probabilities. Nevertheless, we lacked the tools to prove that there are exactly solvable Hamiltonians for chains of different lengths and that there is a precise way for obtaining longer and longer exactly solvable chains starting from shorter ones. Using the results of Section~\ref{sec:exact-and-general}, we calculate the whole spectrum of different chain's Hamiltonians. We group different chain lengths in sequences. The length of the chains on a sequence depends on two natural numbers, $k_0$ and $N_0$. $N_0$ is the shortest length of a chain belonging to a sequence, and $N_0=3 k_0 + 5$. Obtaining the other elements of a given sequence implies the glueing technique. So, for instance,  
$N_0=(5, 11, 23, \ldots)$, $N_0=(8, 17, 35,\ldots)$, and $N_0=(14, 29, 59,\ldots)$,  originate different sequences.


In Table 1, we tabulate the eigenvalues of the one-excitation Hamiltonians for homogeneous qubit chains with the architecture shown in the cartoons of Figures 1 and 2. From top to bottom, the Table shows the new eigenvalues that appear for the chain with length $2N+1$ since the eigenvalues of the Hamiltonian of the chain with length $N$ are also eigenvalues of the Hamiltonian of the chain with length $2N+1$, as shown in Lemma 4. 

\begin{table}{h}
\begin{tabular}{c|c|c|c}
     & $N_0 = 5$ & $N_0 = 8$ & $N_0 = 14$ \\ \hline
     $N_0$ &  1 : $\sqrt{3}$ &  1 : 2 : $\sqrt{2}$ & 1 : $\sqrt{\left( 5 \pm \sqrt{5}
     \right) / 2} $ : $\sqrt{\left( 7 \pm \sqrt{5} \right) / 2} $\\
     $2 N_0 + 1$ & $\sqrt{3 \pm \sqrt{2}}$ & $\sqrt{3}$ : $\sqrt{3 \pm \sqrt{3}}$ &
     $\sqrt{3}$ : $\sqrt{3 \pm \sqrt{\left( 5 \pm \sqrt{5} \right) / 2}}$\\
     $4 N_0 + 3$ & $\sqrt{3 \pm \sqrt{2 \pm \sqrt{2}}}$ & $\sqrt{3 \pm \sqrt{2}}$ :
     $\sqrt{3 \pm \sqrt{2 \pm \sqrt{3}}}$ & $\sqrt{3 \pm \sqrt{2}}$ : $\sqrt{3 \pm
     \sqrt{2 \pm \sqrt{\left( 5 \pm \sqrt{5} \right) / 2}}}$\\
     $8 N_0 + 7$ & $\sqrt{3 \pm \sqrt{2 \pm \sqrt{2 \pm \sqrt{2}}}}$ & $\sqrt{3 \pm
     \sqrt{2 \pm \sqrt{2}}}$ : $\sqrt{3 \pm \sqrt{2 \pm \sqrt{2 \pm \sqrt{3}}}}$
     & 
   \end{tabular}
   \caption{\label{tab:tabla1} The Table tabulates the exact positive eigenvalues of qubit chains with all the couplings equal to the unity. As qubit chains with all their couplings equal are symmetrical, all the lemmas in Section \ref{sec:exact-and-general} hold and, in particular, can be used to construct even larger bi-symmetric chains. The spectrum of a bi-symmetric chain with $2N+1$ qubits contains the spectrum of the symmetric chain with $N$ qubits, so we only show the non-zero eigenvalues of the chain with $2N+1$ qubits different from those tabulated on the row of the chain with $N$ qubits. The whole spectrum of a chain with a given length is obtained using the data in Table and the lemmas on the previous Section. For details see the text.  }
\end{table}

The spectrum for the one excitation Hamiltonian of a chain with a particular length has as many eigenvalues as qubits there are in the chain. For instance, the spectrum of a chain of $N=29$ qubits contains all the eigenvalues on the first two files of the fourth column in Table 1, which are 10. Then, accordingly with the lemmas in Section \ref{sec:exact-and-general}, the spectrum also contains the negatives of these first ten eigenvalues, making the total equal to 20. After that, to complete the spectrum, add nine null eigenvalues to the previous 20 since the one excitation Hamiltonian has $k+1$ null eigenvalues where $k=8$ satisfies $3k+5=29$. 

It is relatively simple to identify the homogeneous chains whose Hamiltonians have a spectrum with exact algebraic expressions for their eigenvalues using arguments similar to those employed in the previous paragraphs. Appendix A includes the positive eigenvalues for the homogeneous chain with $44$ qubits,  the shortest chain of a sequence. We also include a Table that contains information about the characteristics of the spectrum of chains with length $3k+5$ and $1\leq k \leq 100$. There, and in the following, we use $S_{N_0}$ to denote the sequence of chains whose shortest element has a length of $N_0$ qubits. For instance, the sequence originated by $N_0=5$ has elements $S_5= \lbrace 5, 11, 23, 47, \ldots\rbrace$. 

We look for the transmission probability between the corner qubits on a given chain. The transmission probability between the corner qubits for a chain with 17 qubits is given by

\begin{eqnarray} \label{epN17g1}
P_{17} = \frac{1}{144} \left[ -2 \cos (t) - \cos (2t) - 3 \cos \left( 
\sqrt{2} \, t \right) + 2 \cos \left( \sqrt{3}\, t \right) \right. + \nonumber \\
 \left. (2+\sqrt{3}) cos \left( \sqrt{3-\sqrt{3}}\, t \right) +
(2-\sqrt{3}) cos \left( \sqrt{3+\sqrt{3}}\, t \right) .
\right]^2
\end{eqnarray}

To obtain the transmission probability in the equation above, we use the eigenvalues tabulated in Table 1, in particular, those in the column corresponding to $N_0=8$ since $S_8=\lbrace 8,17, 35, \ldots\rbrace$. Note that the cosine functions have in their arguments the values $1, 2, \sqrt{2}, \sqrt{3}$, and $\sqrt{3\pm\sqrt{3}}$ that are the tabulated eigenvalues. 

Showing that a transmission probability like the one in Eq.~\eqref{epN17g1} is compatible with pretty good transmission depends on the observation that PGT implies that $P_{17}$ should become arbitrarily close to the unity. For this to happen,  all the cosine function values simultaneously must be close to $\pm 1$. In this case, all the coefficients multiplying the cosine functions add up to 12 and the condition that $P_{17}\simeq 1$ is satisfied. Using a well-known theorem of Dirichlet dealing with rational approximations of irrational numbers \cite{Dirichlet}, we showed that transmission probabilities like the one in Eq.~\eqref{epN17g1} are compatible with PGT and that the arrival time where it becomes closer to the unity scale with the number of linearly independent irrational numbers contained in the spectrum of the one excitation Hamiltonian of the qubit chain \cite{Serra2024}.

For longer and longer chains, the expression for the exact transmission probabilities becomes more and more cumbersome to handle. Nevertheless, the data showed in Table II (see  Appendix A) show that the $S_5$, $S_8$, $S_{14}$, and $S_{44}$ sequences have exact algebraic equations determining their spectra. 

\section{Transmon qubit chains with perfect state transfer}
\label{sec:perfect}

To find chains whose Hamiltonians have exact analytical and algebraic spectra compatible with perfect transmission, we employ again the lemmas stated in Section \ref{sec:exact-and-general}. We aim to solve the inverse problem. By prescribing eigenvalues compatible with perfect transmission, we find the one-excitation Hamiltonian having the prescribed spectrum. The solution of the inverse problem allowed the design of XX linear chains showing perfect quantum state transmission. The different architecture of the chains considered in this work precludes the same procedure because the spectra of symmetric transmon chains, as those depicted in Figure 2, always have several null eigenvalues, and the Hamiltonian matrix is not tridiagonal. We will illustrate our algorithm by solving the two shortest non-trivial chains, $N=8$ and $11$.

 Nevertheless, before presenting the exact results, it is worth mentioning that a matrix with the structure shown by the Hamiltonian matrix in Eq.~\eqref{ec:h8symm} is called a symmetric doubly bordered band diagonal matrix. This matrix type often appears in inverse eigenvalue problems, and numerous methods exist to efficiently tridiagonalize them. The exact analytical tridiagonalization is quite cumbersome to write down. Because of this, we explicitly present only a pair of cases.
 

\subsection*{$N=8$ case}
As can be appreciated in Figure 2, the chain with $N=8$ qubits has four different interaction coefficients, $v_1$, $v_2$, $g_1$, and $g_2$. Accordingly, with the lemmas in Section \ref{sec:exact-and-general}, the spectrum of $h_8^s$ has $2$ null eigenvalues, and the other $6$  come in pairs $\pm \lambda$. Choosing $\pi$ as the arrival time for the perfect quantum state transfer imposes that the eigenvalues must be integer numbers alternating between odd and even values. For simplicity, we choose consecutive integer eigenvalues as follows
\begin{equation}
E_{\pm 1} = \pm k, E_{\pm} = \pm (k\pm 1), E_{\pm 2} = \pm (k+2), E_0 = 0 \mbox{(deg. 2)} . 
\end{equation}

The characteristic polynomial for the eigenvalues factorizes in three equations for the four interaction coefficients. Solving for $g_1$, we get that
\begin{equation}
g_1^2=k-v_1^2 , \quad g_1^2=(k+1)^2 - v_1^2, \quad  (k+2)^2 - v_1^2.
\end{equation}

Choosing from the three solutions the one in the middle, to ensure  $v_2>0$, we get
\begin{equation}
v_2^2 = \frac{3+4k(k+2)}{2v_1^2} \quad g_1^2 = \frac{  3 (v_1^2 - 1) - k  (k+2) (4-v_1^2) }{v_1^2} , 
\end{equation}
Using both Equations above, and asking that $g_1^2>0$, and that $g_2^2>0$ , we arrive to the condition

\begin{equation}
\frac{4k^2 + 8 k +3}{k^2 + 2k + 3} < v_1^2 < (k+1)^2 . 
\end{equation}

Putting all the eigenvalues and eigenvector of the one excitation Hamiltonian in terms of $k$, The transmission probability between the corner qubits is given by
\begin{equation}\label{ec:P8}
P(t) = \frac{1}{64 (k+1)^2} \left(
(2k+3) \cos(k t) - 4(k+1) \cos((k+1)t) + (2k+1) \cos((k+2) t )
\right)^2 ,
\end{equation}
that satisfies $P((2\ell+1)\pi)=1$, for all $\ell \in \mathbb{Z}$. 




\subsection*{$N=11$ case}

Assuming that the eigenvalues of the one-excitation Hamiltonian matrix are  $E_{\pm 1}= (k\pm 1)$, $E_{\pm 2}= (k\pm 2)$, $E_{\pm 3}= (k\pm 3)$, $E_{\pm 4}= (k\pm 4)$, and $E_0=0$ has threefold degeneracy, and solving the inverse eigenvalue equations in terms of $v_1$ and $k$ we get that
\begin{equation}
v_2 = \frac{1}{2v_1} \sqrt{3(5+12 k +4 k^2)}, \quad v_3= \sqrt{3+2k},
\end{equation}
that are acceptable solutions for any value of $k$. For the other interaction coefficients, we get
\begin{equation}
g_1 = \sqrt{\frac{1}2{} (3+6 k +2 k^2 - 2v_1^2)} , \qquad g_2 = \frac{1}{2 v_1} \sqrt{(10+4k+4k^2) v_1^2 - (15+36k+12 k^2)} . 
\end{equation}
Demanding that $g_1^2>0$ and  $g_2^2>0$ imposes that $v_1^2$ satisfies that
\begin{equation}
  \frac32 \frac{4k^2+12 k +5}{2k^2 + 2k + 5}< v_1^2  <\frac12 (2k^2+6k+3)^2. 
\end{equation}
By choosing $v_1=2$, which is the only natural number that satisfies the condition, we get that the transmission probability between the corner qubits is given by

\begin{eqnarray}\label{ec:P11}
P_{11} &=& \frac{1}{2^8((k+1)(k+2))^2} \left(((10+9k+2k^2)\cos (kt) - 3 (5+7k+2k^2) \cos ((k+1)t) + \right. \nonumber \\ 
&&\left. (1+2k) (3(2+k) \cos ((2+k)t) - (1+k) \cos ((3+k)) t )  \right)^2 ,
\end{eqnarray}
which satisfies that $P_{11}(t=(2\ell+1)\pi) )=1$, and $P_{11}(t=(2\ell)\pi) )=0$, for all $\ell\in \mathbb{Z}$.

\hspace{1ex}

\section{An upper bound for the probability of transmission for dimmerized chains}
\label{sec:upper}

The use of dimerized transmon qubit chains, or more generally, dimerized chains showing topological states as good quantum state transfer channels, has been suggested in many papers. Nevertheless, we aim to show that this scenario does not appear in practical times. Using exact and numerical results, we show that the transmission probability of homogeneous chains is an upper bound of the transmission probabilities of dimerized chains, irrespective of the dimerization strength and the value of the interaction coefficient $g$. 

We proceed as follows. First, we obtain the exact transmission probability between the corner qubits of a chain with the Hamiltonian in Eq.~\eqref{ec:ssh-hamiltonian} such that $t_i=1$, $w_i=w$, and $g_i=g$, for all $i$. We will consider here, as an example, a chain with $N=11$ qubits. The transmission probability is a function of two parameters, $w$ and $g$. As shown in this paper and previous ones \cite{Serra2022JPA,Serra2024}, the transmission probability reaches its maximum value when all the cosine functions in its expression are equal to $\pm 1$ (see Equations \eqref{epN17g1}, \eqref{ec:P8}, and  \eqref{ec:P11}). For fixed values of $w$, we find the optimal value of $g$ for which the transmission probability attains its maximum.

The exact transmission probability for a chain with $N=11$ qubits is as follows.

\begin{eqnarray}\label{ec:probabilidad-exacta-dim}
P(t=1,g_i=g,w_i=w;t)\,=\,\left[
\frac{w}{4 \left(w^2+1\right)
   \left(w^4+1\right)} \left(\left(w^2+1\right) \left(\left(w
   \left(w+\sqrt{2}\right)+1\right) \cos \left(\sqrt{g^2+w^2-\sqrt{2}
   w+1} \,t \right)+ \right. \right. \right.  \nonumber \\
\left. \left. \left.
\left(w^2-\sqrt{2} w+1\right) \cos \left(\sqrt{g^2+w
   \left(w+\sqrt{2}\right)+1}\, t\right)\right)-2 \left(w^4+1\right) \cos \left(
   \sqrt{g^2+w^2+1} \, t\right)-4 w^2 \cos (g \, t)\right)\right]^2 .
\end{eqnarray}
Implementing the replacement of the cosine functions and optimizing the resulting expression, we get that the upper bound for the transmission probability is given by

\begin{equation}\label{ec:probabilidad-max-opt}
P_{up}=\left(\frac{w\,(1+w^2)}{1+w^4}\right)^2 \,.
\end{equation}

\begin{figure}[h]
\includegraphics[width=0.7\linewidth]{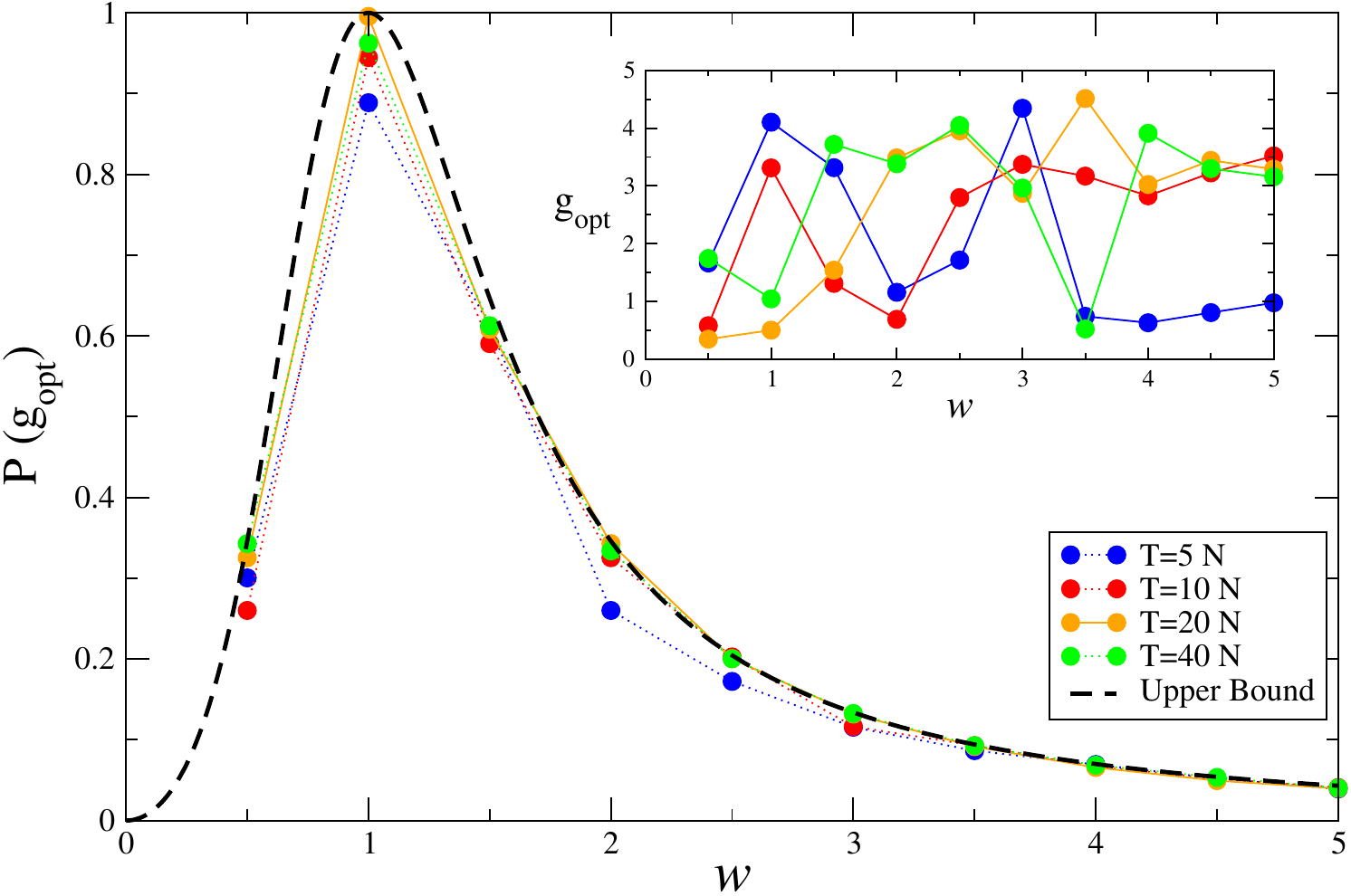}
\caption{The Figure shows the behaviour of $P_{up}$, Eq.~\eqref{ec:probabilidad-max-opt},  {\em vs} the dimerization parameter $w$ for a chain with $N=11$ qubits. Besides the values of $P_{up}$, we plotted the numerical values obtained for the transmission probability using a global optimization algorithm that finds the value of $g$ with the maximum transmission probability. The optimization algorithm looks for maximum values of the transmission probability at fixed, chosen transmission times and values of the dimerization parameter. The blue, red, yellow and green solid dots correspond to transmission times $T=5N, 10N, 20N$, and $40N$ and show the values of the transmission probability found using the optimization algorithm. It is clear that the transmission probability is an increasing function of the transmission time and that, for relatively short times, the optimization values come close to the exact results, especially for large values of $w$. More importantly, the results in the Figure show that dimerization does not provide chains with good transfer abilities when compared with homogeneous ones. The inset shows the corresponding values of $g$ obtained by the optimization algorithm for each pair of values of $w$ and $T$.    }
\label{fig:upper-bound}
\end{figure}

In Figure~\ref{fig:upper-bound}, we plot the maximum value of transmission probability as a function of $w$ at the optimal value of $g$. The black dashed curve corresponds to the maximum value of the transmission probability. The curve shows a clear maximum for $w=1$, i.e. the homogeneous chain is the best transfer channel. Of course, this finding produces at least two questions. Is this upper-bound achievable at short arrival times? For larger chains where an analytical treatment becomes unfeasible, is there a way to detect if the dimerized chain offers an advantage over homogeneous ones as transfer channels? 

To answer the first question, we use an algorithm implementing a global optimization method to determine the optimal value of the transmission probability and the corresponding parameters determining the optimal value\cite{serraGO}. In Figure~\ref{fig:upper-bound}, we include the values found fixing the values of $w$ and the arrival time $T$, so the optimization method only searches on a one-dimensional space. We have used this optimization method when looking for very high values of the transmission probability in Heisenberg-like chains and considering all the exchange coupling interactions as variables. In this case, the algorithm searches in a $N$ dimensional space where $N$ is the length of the chain. In the following, we will use the algorithm in search spaces with few variables.

As can be appreciated in Figure~\ref{fig:upper-bound}, the analytical upper bound value lies above the values found with the optimization method, which correspond to the coloured solid dots. The values for the transmission probability found with the optimization methods are increasing functions of the arrival time. For arrival times on the order of $20N$ or $40N$, the numerical probability values are indistinguishable from the analytical ones, even for low values of $w/v$, the dimerization parameter.

\section{Transmon chains with optimized couplings}
\label{sec:optimized-couplings}

The transmission probability plotted in Figure~\ref{fig:upper-bound} is pretty conclusive about the possibility of using dimerized chains to transmit quantum states with high fidelity at practical times, i.e., the homogeneous chain without dimerization is a far superior channel to send quantum states. Nevertheless, since the eigenstates whose energy lies within the energy gap that appears when $w/v\neq 1$ and large enough are protected,  in the sense that their energy and the localization of their eigenvectors vary less when the chain is disturbed than the extended states whose energies are in the energy bands typical of an XX chain model it is tempting looking for dimerized chains with short transmission times.  

A possible way out of this problem, i.e., the poor transmission probabilities characteristic of dimerized chains in which $t$, $w$, and $g$ have constant values, consists of keeping the dimerization but using the g-qubits as controls whose interaction strength can be optimized site-by-site to obtain a better transmission channel than the one with all the values $g_i$ equal. In this way, we would improve transmission while maintaining the highly valued properties of topological chains. Along this Section, we explore this idea.

\begin{figure}[h]
\includegraphics[width=0.6\linewidth]{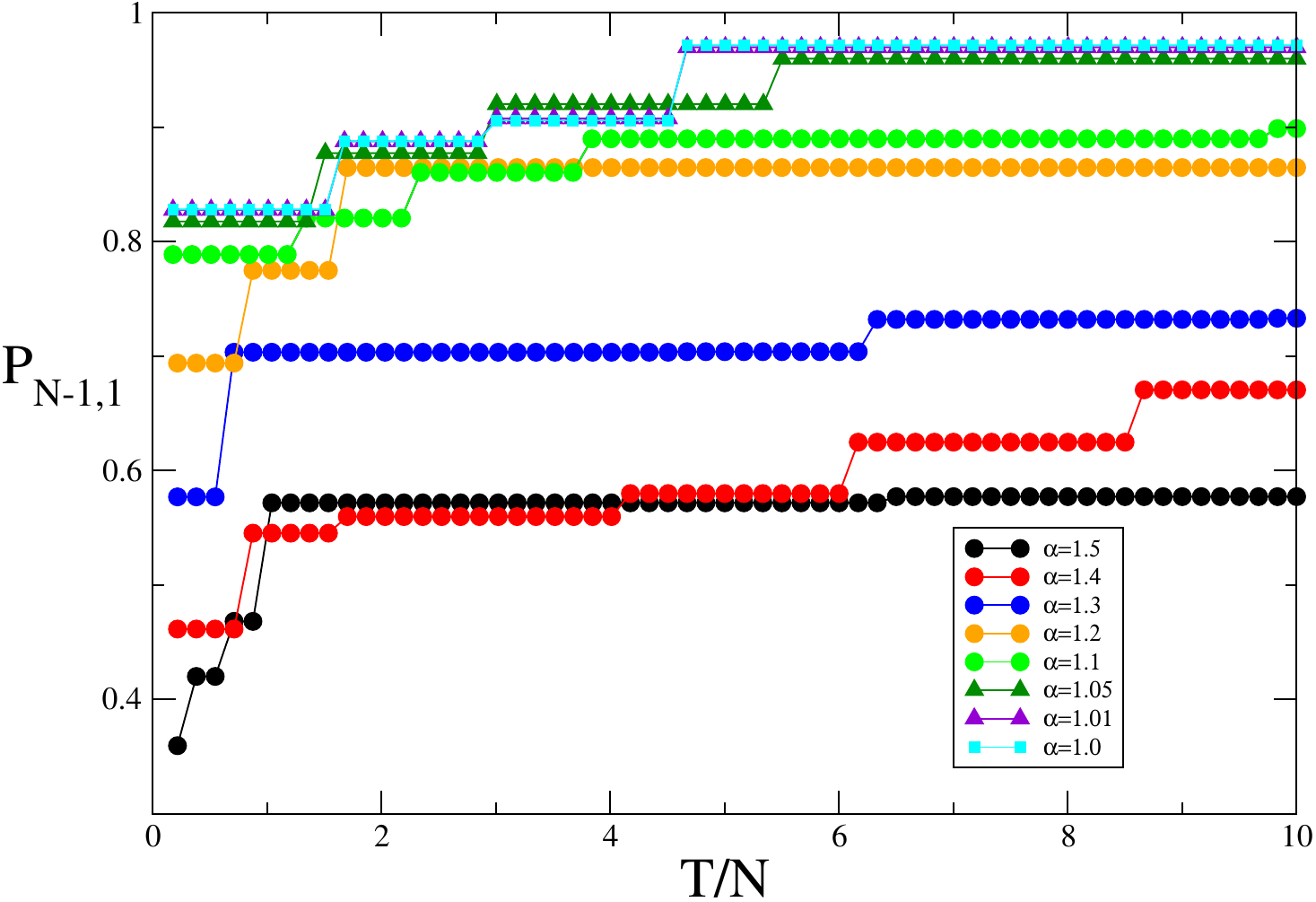}
\caption{The Figure shows the transmission probability between the corner qubits for a chain with $14$ qubits. We obtain the probability by running the global optimization algorithm with only two variables, $t$ and $g$. The variable $w$ satisfies that $t = \alpha w$. Each coloured point corresponds to an arrival time $T$, a fixed quantity necessary for the algorithm, and the same colour identifies all the points corresponding to a single value of $alpha$, as shown in the legend of the figure. For $alpha=const.$ the probability is an increasing function of the arrival time, growing through discreet steps until it reaches a maximum value. This maximum value depends on $\alpha$, which is larger for $\alpha=1$.   }
\label{fig:otro-bound}
\end{figure}

\begin{figure}[h]
\includegraphics[width=0.6\linewidth]{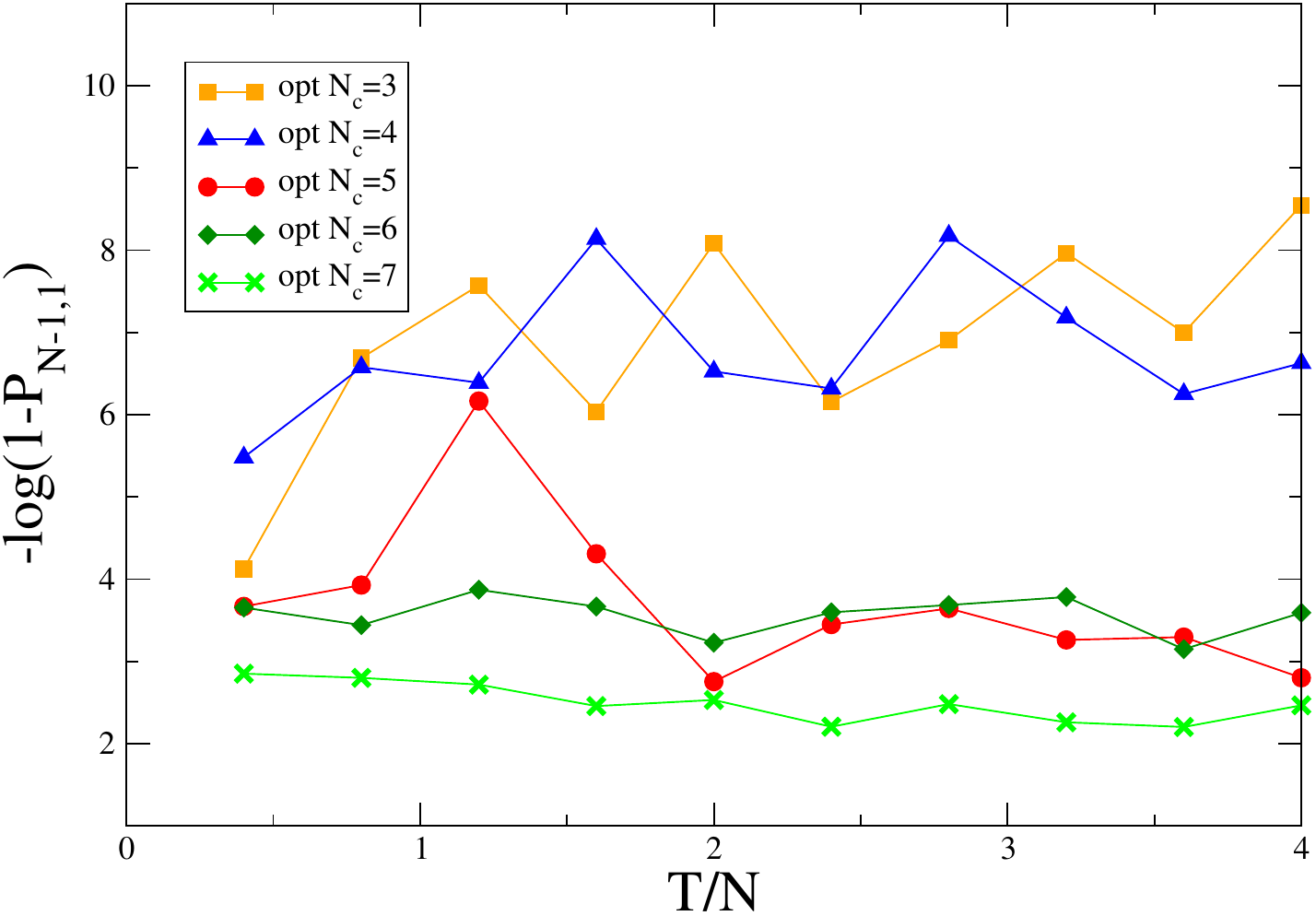}
\caption{The Figure shows the transmission probability obtained for chains of different numbers of qubits, $3k+5$, where $k=2, 3, 4, 5$, and $6$ and different transmission times. Since all the probability values lie between 0.99 and 1, we plot the values of $-\log (1-P)$ instead of the probability $P$ to widen the differences between the data. The solid dots correspond to the numerical results, and the lines are a guide to the eye. The yellow, blue, red, dark green and light green solid dots correspond to the transmission probabilities of the chains with  $k=2, 3, 4, 5$, and $6$, respectively. The optimization procedure results in chains having excellent transfer abilities, with probabilities bigger than $0.99$ in all the cases shown.   }
\label{fig:cuantos-nueves}
\end{figure}

There are two ways to explore. In the first, the ratio between $t$ and $w$ remains constant, and the optimization variables are the strength of the g-couplings and the strength of $t$. In the second, $t$ and $w$ are independent variables in the same fashion as the strength of the g-couplings. Figure \ref{fig:otro-bound} shows the results obtained exploring the first alternative, while Figure \ref{fig:cuantos-nueves}  shows the results of the second alternative.

In the previous Section, we explored what kind of results we obtained by running the optimization algorithm keeping the ratio $t/w$ and $g$ constant, but only included results for large values of the transmission time. Figure \ref{fig:otro-bound} shows the transmission probability obtained for a chain with $14$ qubits for several values of $t/w$ and short transmission times.It is clear that the transmission probability is an increasing function of the transmission time, but even for the chain with $t=w=1$ short transmission times ($T\leq 4N$) implies that $P\leq 0.9$. 

Running the optimization algorithm with $t$, $w$, and all the g-qubits as optimization variables gives us $k+4$ variables to optimize for a chain with $N=3k+5$ qubits. Figure \ref{fig:cuantos-nueves} shows the results for the transmission probability obtained by optimizing the $k+4$ variables for five values of $k$. Note that the vertical axis corresponds to $-\log{(1-P_{3k+5})}$ since all the transmission probabilities lie above $0.99$. This function magnifies the minimal differences between the transmission probabilities found for chains with different lengths. It is clear that $-\log{(1-0.99)}=2$, $-\log{(1-0.999)}=3$, and so on.  We scale the arrival times using the chain length, which is one of the main parameters of the problem, being the other the maximum value of the interaction coefficients. For short chains, $k\leq 6$, the transmission probability is larger than $0.996$, which is more than enough for practical applications. For larger and larger chains, the performance gradually worsens. 

These results show that dimerized chains controlled through the g-qubits easily outperform the dimerized chains with $g_i=g$, $t_i=t$ and $w_i=w$ for short arrival times.

\section{Discussion and Conclusions}
\label{sec:conclusions}

We have presented two families of exact solutions that result in different transmission scenarios whose implications for actual quantum state transmission in decorated transmon chains are evident, at least if a fast, high-quality transmission is required. Homogeneous chains with all their coupling coefficients equal only support pretty good quantum state transfer, with exceedingly long transmission times. Chains with site-dependent couplings meet the requirements of fast and perfect quantum state transmission by tuning the values of the couplings accordingly with the exact solution presented. 

The upper bound provided by the exact solution of a short dimerized chain  (Section~\ref{sec:upper}), on the other hand, provides clear evidence that uncontrolled dimerized chains do not offer better quantum state transfer than homogeneous ones. This assessment holds for larger chains. We do not include explicit expressions for longer chains for brevity since the exact expressions of eigenvalues and eigenvectors occupy a few pages. 

The sequences of exactly solvable chains introduced in Section~\ref{sec:pretty-good} have eigenvalues with algebraic expressions, {\em i. e.} the characteristic polynomial of their Hamiltonian matrices is exactly factorizable as a product of second-degree polynomials. The chains whose lengths do not belong to a sequence have characteristic polynomials that do not factorize in a product of second-degree polynomials or less. In Table~\ref{tab:sequences}, we provide the degree of the polynomials determining the eigenvalues of the Hamiltonian matrices after the factorization of the characteristic polynomial for decorated chains with up to three hundred qubits. When factoring the characteristic polynomial of chains that do not belong to any sequence, the characteristic polynomial keeps a factor that is a high-degree polynomial (see Table \ref{tab:sequences}). On the other hand, we can not ascertain if there are sequences $S_{N_0}$ that start with $N_0>44$. 

We found the chains that show perfect quantum state transfer using the known properties of their spectra (as stated on the lemmas in Section \ref{sec:exact-and-general}) and assuming, a simple relationship between consecutive eigenvalues, that they are consecutive integer numbers. After this assumption, the inverse problem, determining the coupling strengths, becomes a series of consistency equations that must be satisfied to obtain all the strengths. The inverse problem in XX chains without external fields has an exact solution, {\em i. e.} given the spectrum,  a set of equations determines all the coupling strengths values. So far, we have not found if a similar set of equations, not inequalities, can be written down for a chain or arbitrary length. 

The results of the quantum state properties on chains with optimized couplings are conclusive. A high ratio between $w$ and $t$ is incompatible with fast, high-quality quantum state transfer. The best alternative to achieve this kind of properties for quantum transfer in uncontrolled chains seems to be the construction of chains with $k+4$ optimizable variables, $t$, $w$, and the g-couplings. Searching maximum values in a $k+4$ is not too expensive, even for moderate-length chains. 

The glueing mechanism used to obtain chain sequences exhibiting PGT in Section IV suggests that it is possible to obtain chain sequences with perfect transmission between the corner qubits. However, the analytical effort required is enormous and does not seem generalizable, as in the case of strings exhibiting PGT. We are looking for other qubit layouts showing perfect quantum state transmission for growing distances between the sender and receiver qubits with Hamiltonians compatible with transmon qubits.

\acknowledgments

The authors acknowledge partial financial support from CONICET (PIP 11220210100787CO,
and PIP11220200100170). A F acknowledges partial financial support from ANPCyT (PICT2019-0654) and 
UNNE. O O and P S acknowledge partial financial support
from CONICET and SECYT-UNC.

\appendix

\section{Numbering of the sites of a decorated chain}
\label{ap:numb}

\begin{figure}[h]
\includegraphics[width=0.7\linewidth]{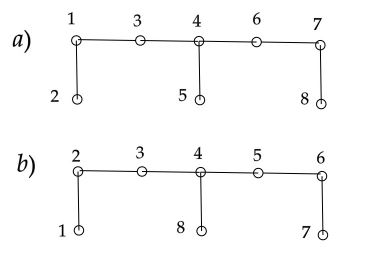}
\caption{The structure of the one-excitation Hamiltonian matrix depends on the numbering assigned to the qubits. The numbering does not change its spectrum but renders some proofs in Section \ref{sec:exact-and-general} simpler. The cartoon shows two numberings, which result in different structures for the one excitation Hamiltonian matrix of a chain with eight qubits. Eqs.~\eqref{ec:h8first} and \eqref{ec:h8symm} show the respective one excitation Hamiltonian matrices. We will return to the numbering subject in the next Section, too.    }
\label{fig:numbering}
\end{figure}

In other studies on the transfer properties of superconducting qubit chains \cite{Serra2024, Wang2022}, the preferred numbering of the qubits can be appreciated in Figure~\ref{fig:numbering} a). On the other hand, in some sections of the present work, we prefer the numbering shown in Figure~\ref{fig:numbering} b). 

For instance, the one excitation Hamiltonian matrix for a chain with $N=8$ qubits, using the numbering in Figure~\ref{fig:numbering}a), is given by
\begin{equation}\label{ec:h8first}
h_8^c = \left(\begin{array}{cccccccc}
     0 & g_1 & v_1 & 0 & 0 & 0 & 0 & 0\\
     g_1 & 0 & 0 & 0 & 0 & 0 & 0 & 0\\
     v_1 & 0 & 0 & w_1 & 0 & 0 & 0 & 0\\
     0 & 0 & w_1 & 0 & g_2 & v_2 & 0 & 0\\
     0 & 0 & 0 & g_2 & 0 & 0 & 0 & 0\\
     0 & 0 & 0 & v_2 & 0 & 0 & w_2 & 0\\
     0 & 0 & 0 & 0 & 0 & w_2 & 0 & g_3\\
     0 & 0 & 0 & 0 & 0 & 0 & g_3 & 0
   \end{array}\right),
\end{equation}
where the interaction coefficients are those in Eq.~(\eqref{ec:ssh-hamiltonian}). On the other hand, using the numbering in Figure~\ref{fig:numbering}b) we get that the corresponding matrix for a symmetric chain with eight qubits is given by
\begin{equation}\label{ec:h8symm}
h_8^s = \left(\begin{array}{cccccccc}
     0 & g_1 & 0 & 0 & 0 & 0 & 0 & 0\\
     g_1 & 0 & v_1 & 0 & 0 & 0 & 0 & 0\\
     0 & v_1 & 0 & v_2 & 0 & 0 & 0 & 0\\
     0 & 0 & v_2 & 0 & v_2 & 0 & 0 & g_2\\
     0 & 0 & 0 & v_2 & 0 & v_1 & 0 & 0\\
     0 & 0 & 0 & 0 & v_1 & 0 & g_1 & 0\\
     0 & 0 & 0 & 0 & 0 & g_1 & 0 & 0\\
     0 & 0 & 0 & g_2 & 0 & 0 & 0 & 0
   \end{array}\right).
\end{equation}
The superscripts in both Hamiltonians correspond to cell and symmetric, emphasising the numbering that produces the specific shape of the Hamiltonian matrices. 

\section{Some results about sequences of exactly solvable chains}
\label{ap:appendix-a}

In Section \ref{sec:pretty-good}, we introduced the sequences of qubit chains that have their spectra formed by exact eigenvalues that are solutions of quadratic equations. In Table \ref{tab:tabla1}, we explicitly showed the eigenvalues of the sequences $S_5, S_8$, and $S_{14}$. Here, we present the explicit eigenvalues of the chain with $44$ qubits 
\begin{equation}\label{ec:eigen-44}
1\quad ; \quad 2 : \quad \sqrt{2} \quad ; \quad \sqrt{(5 \pm \sqrt{5})/2} \quad ; \quad \sqrt{(7 \pm \sqrt{5})/2} \quad ; \quad
\frac{1}{2}  \sqrt{ 13 \pm \sqrt{5} \pm  \sqrt{30\pm 6\sqrt{5}} } . 
\end{equation}

Counting the number of eigenvalues is instructive. Since each $\pm$  sign represents "two possibilities", then there are $2^3=8$ eigenvalues in the expression
\begin{equation}
\frac{1}{2}  \sqrt{ 13 \pm \sqrt{5} \pm  \sqrt{30\pm 6\sqrt{5}} } . 
\end{equation}
So, from left to right, there are 1, 1, 1, 2, 2, and $8$ eigenvalues in Eq.~\eqref{ec:eigen-44}. Summing up, there are $15$ definite positive eigenvalues in the spectrum, so $2 \times 15 + 14=44$, where we use that the chain with $44$ qubits has $k+1$ null eigenvalues, and $k$ satisfies that $3k+5=44$.

\begin{table}[ht]
  \begin{tabular}{c|c|c|c||c|c|c|c|| c|c|c|c}
    k & N & Sequence & poly & k & N & Sequence & poly  & k & N & Sequence & poly \\ \hline
    1 & 8 & $S_8$ & 2     & 35 & 110 &  & 3         & 69 & 212 &  & 35\\
    2 & 11 & $S_5$ & 2    &  36 & 113 &  & 11       & 70 & 215 &  & 3\\
    3 & 14 & $S_{14}$ & 2 & 37 & 116 &  & 23        & 71 & 218 &  & 36\\
    4 & 17 & $S_8$ & 2    & 38 & 119 & $S_{14}$ & 2 & 72 & 221 &  & 18\\
    5 & 20 &  & 3         &  39 & 122 &  & 20       & 73 & 224 &  & 20\\
    6 & 23 & $S_5$ & 2    &  40 & 125 &  & 6        & 74 & 227 &  & 9\\
    7 & 26 &  & 3         &  41 & 128 &  & 21       & 75 & 230 &  & 30\\
    8 & 29 & $S_{14}$ & 2 &  42 & 131 &  & 5        & 76 & 233 &  & 12\\
    9 & 32 &  & 5         &  43 & 134 &  & 3        & 77 & 236 &  & 39\\
    10 & 35 &  & 2        &  44 & 137 &  & 11       & 78 & 239 & $S_{14}$ & 2\\
    11 & 38 &  & 6        &  45 & 140 &  & 23       & 79 & 242 &  & 3\\
    12 & 41 &  & 3        &  46 & 143 & $S_8$ & 2   & 80 & 245 &  & 20\\
    13 & 44 & $S_{44}$ & 2 & 47 & 146 &  & 21       & 81 & 248 &  & 6\\
    14 & 47 & $S_5$ & 2   &  48 & 149 &  & 10       & 82 & 251 &  & 6\\
    15 & 50 &  & 8        &  49 & 152 &  & 16       & 83 & 254 &  & 32\\
    16 & 53 &  & 3        &  50 & 155 &  & 6        & 84 & 257 &  & 21\\
    17 & 56 &  & 9        &  51 & 158 &  & 26       & 85 & 260 &  & 28\\
    18 & 59 & $S_{14}$ & 2 &  52 & 161 &  & 3       & 86 & 263 &  & 5\\
    19 & 62 &  & 6        &  53 & 164 &  & 20       & 87 & 266 &  & 44\\
    20 & 65 &  & 5        &  54 & 167 &  & 3        & 88 & 269 &  & 3\\
    21 & 68 &  & 11       &  55 & 170 &  & 18       & 89 & 272 &  & 36\\
    22 & 71 & $S_8$ & 2   &  56 & 173 &  & 14       & 90 & 275 &  & 11\\
    23 & 74 &  & 10       &  57 & 176 &  & 29       & 91 & 278 &  & 30\\
    24 & 77 &  & 6        &  58 & 179 & $S_{44}$ & 2 & 92 & 281 &  & 23\\
    25 & 80 &  & 3        &  59 & 182 &  & 14       & 93 & 284 &  & 36\\
    26 & 83 &  & 3        &   60 & 185 &  & 15      & 94 & 287 & $S_8$ & 2\\
    27 & 86 &  & 14       &  61 & 188 &  & 6        & 95 & 290 &  & 48\\
    28 & 89 & $S_{44}$ & 2 & 62 & 191 & $S_5$ & 2   & 96 & 293 &  & 21\\
    29 & 92 &  & 15       &  63 & 194 &  & 24       &  97 & 296 &  & 10\\
    30 & 95 & $S_5$ & 2   &   64 & 197 &  & 10      & 98 & 299 &  & 10\\
    31 & 98 &  & 20       &  65 & 200 &  & 33       & 99 & 302 &  & 50\\
    32 & 101 &  & 6       &  66 & 203 &  & 8        & 100 & 305 &  & 16 \\
    33 & 104 &  & 21      &  67 & 206 &  & 22       &     &     &   &    \\
    34 & 107 &  & 5       &  68 & 209 &  & 12       &
  \end{tabular}
  \caption{\label{tab:sequences} The Table shows how the characteristic polynomial of the Hamiltonian matrix of a chain factorizes. If the characteristic polynomial factorizes as a product of second-degree polynomials, the whole spectrum is given by exact algebraic expressions. For $k=0, 1, 2, \ldots 100$, and chain lengths $3k+5$, the Table shows if a chain belongs to a given sequence $S_{N_0}$ of bi-symmetric chains, and the degree of the polynomial with the higher degree in which the characteristic polynomial factorizes. Check that for the chains in a sequence, the degree is just two. The characteristic polynomial for chains with lengths beyond $305$ qubits becomes unmanageable. }
\end{table}

The degree of the characteristic polynomial of a Hamiltonian matrix is equal to the number of its columns (or rows). In the one excitation subspace, the Hamiltonian matrix of a chain with $N$ qubits has $N\times N$ dimension and $N$ eigenvalues. The eigenvalues will have explicit expressions, in terms of square roots, only if the characteristic polynomial factorizes as a product of second-degree polynomials or less. This factorization is not always possible. Depending on the length of the chain, unfactorizable high-degree polynomials prevent the finding of explicit expressions for a set of eigenvalues. Table ~\ref{tab:sequences} includes the degree of the polynomial with the highest degree after extracting all the exact polynomials of a degree less or equal to two from the characteristic polynomial.


\begin{thebibliography}{90}

\bibitem{Chang2023} A. Chang and H. Zhan, J. Phys. A: Math. Theor. {\bf 56}, 165305 (2023).

\bibitem{Keele2022} C. Keele and A. Kay, Phys. Rev. A {\bf 105}, 032612 (2022).

\bibitem{Xie2023} W. Xie, A. Kay, and C. Tamon, Phys. Rev. A {\bf 108}, 012408 (2023).

\bibitem{Maleki201} Y. Maleki , A. M. Zheltikov, Optics Communications {\bf 496}, 126870 (2021).

\bibitem{Serra2022JPA} P. Serra, A. Ferrón and O. Osenda, J. Phys. A: Math. Theor. {\bf 55}, 405302 (2022).

\bibitem{Mograby2021} G. Mograby , M. Derevyagin , G. V. Dunne and A. Teplyaev, J. Phys. A Math. Theor. {\bf 54} 125301 (2021).

\bibitem{Christandl2004} M. Christandl, N. Datta, A. Ekert, and A. J. Landahl, Phys. Rev. Lett. {\bf 92}, 187902 (2004).

\bibitem{Nakamura1999} Y. Nakamura, Y. A. Pashkin, and J. S. Tsai, Nature {\bf 398}, 786 (1999).

\bibitem{Krantz2019} P. Krantz , M. Kjaergaard , F. Yan; T. P. Orlando, S. Gustavsson, and W. D. Oliver , Appl. Phys. Rev. {\bf 6}, 021318 (2019).

\bibitem{Stehlik2021} J. Stehlik, D. M. Zajac, D. L. Underwood, T. Phung, J. Blair, S. Carnevale, D. Klaus1, G. A. Keefe, A. Carniol, M. Kumph, M. Steffen, and O.E. Dial, Phys. Rev. Lett. {\bf 127}, 080505 (2021).

\bibitem{Koch2007} J. Koch, T. M. Yu, J. Gambetta, A. A. Houck, D. I. Schuster, J. Majer, A. Blais, M. H. Devoret, S. M. Girvin, and R. J. Schoelkopf, Phys. Rev. A. {\bf 76},  042319 (2007).

\bibitem{Schereier2008} J. A. Schreier, A. A. Houck, J. Koch, D. I. Schuster, B. R. Johnson, J. M. Chow, J. M. Gambetta, L. Frunzio, M. H. Devoret, S. M. Girvin, and R. J. Schoelkopf, Phys. Rev. B. {\bf 77}, 180502(R) (2008). 

\bibitem{Kim2023} Y. Kim, A. Eddins, S. Anand, K. X. Wei, E. van den Berg, S. Rosenblatt, H. Nayfeh, Y. Wu, M. Zaletel, K. Temme, and  A. Kandala, Nature {\bf 618}, 500 (2023).

\bibitem{GoogleAI2023} Google Quantum AI and Collaborators, Nature {\bf 622}, 481 (2023). 

\bibitem{Wang2022} C. Wang, L. Li, J. Gong, and Y. Liu, Phys. Rev. A {\bf 106} 052411 (2022). 

\bibitem{Estarellas2017} M. P. Estarellas, I. D’Amico, and T. P. Spiller,  Sci. Rep. {\bf 7} 42904 (2017).

\bibitem{Lemonde2018} M-A. Lemonde, V. Peano, P. Rabl, and D. G. Angelakis, New J. Phys. {\bf 21}, 113030 (2019).

\bibitem{Lemonde2019} M-A. Lemonde, S. Meesala, A. Sipahigil, M. J. A. Schuetz, M. D. Lukin, M. Loncar, and P. Rabl, Phys. Rev. Lett. {\bf 120}, 213603 (2018).

\bibitem{Mistakidis2020} S. I. Mistakidis, G. M. Koutentakis, G. C. Katsimiga, T. Busch, and P. Schmelcher, New J. Phys.
{\bf 22} 033030 (2020).

\bibitem{Yousefjani2021a} R. Yousefjani and A. Bayat, Quantum {\bf 5} 460 (2021).

\bibitem{Yousefjani2021b} R. Yousefjani, S. Bose, and A. Bayat, Phys. Rev. Res. {\bf 3} 043142 (2021).

\bibitem{Serra2024} P. Serra, A. Ferrón, and O. Osenda, J. Phys. A: Math. Theor. {\bf 57}, 015304 (2024). 

\bibitem{Burgarth2006} D. Burgarth, Ph.D. thesis, University College London, 2006.

\bibitem{Godsil2012a} C. Godsil, S. Kirkland, S. Severini, and J. Smith, Phys. Rev. Lett. {\bf 109}, 050502 (2012).

\bibitem{Godsil2012b} C. Godsil, Discrete Math. {\bf 312}, 129 (2012).

\bibitem{Vinet2012} L. Vinet and A. Zhedanov, Phys. Rev. A {\bf 86}, 052319 (2012).

\bibitem{Kay2010} A. Kay, Int. J. Quantum Inf. {\bf 8}, 641 (2010).

\bibitem{Banchi2017} L. Banchi, G. Coutinho, C. Godsil, and S. Severini, J. Math. Phys. {\bf 58}, 032202 (2017).

\bibitem{Godall2012} C. Godsil, S.t Kirkland, S. Severini, and J. Smith, Phys. Rev. Lett. {\bf 109}, 050502 (2012).


\bibitem{compu-mathema}Z-y. Peng, X-y. Hu, and L. Zhang, J. of Comp. Math. {\bf 22}, 535 (2004).

\bibitem{Coutinho2016} G. Coutinho, Electron. J. Comb. {\bf 23}, 46 (2016).

\bibitem{6hristandl2005} M. Christandl, N. Datta, T. C. Dorlas, A. Ekert, A. Kay, and A. J. Landahl, Phys. Rev. A. {\bf 71}, 032312 (2005).

\bibitem{Burgarth2005} D. Burgarth, V. Giovannetti, and S. Bose, J. Phys. A: Math. Gen. {\bf 38}, 6793 (2005).

\bibitem{Bayat2014} A. Bayat, Phys. Rev. A {\bf 89}, 062302 (2014).

\bibitem{Banchi2010} L. Banchi, T. J. G. Apollaro, A. Cuccoli, R. Vaia, and P. Verrucchi Phys. Rev. A {\bf 82}, 052321 (2010).

\bibitem{Zwick2015} A. Zwick, G. A. Alvarez, J. Stolze, and O. Osenda, Quantum Information and Computation {\bf 15}, 0582
(2015).

\bibitem{Loft2016} N. J. S. Loft et al New J. Phys. {\bf 18}, 045011 (2016).


\bibitem{Banchi2011} L. Banchi, A. Bayat, P. Verrucchi, and S. Bose, Phys. Rev. Lett. {\bf 106}, 140501 (2011).


\bibitem{Chapman2016} R. J. Chapman, M. Santandrea, Zixin Huang, G. Corrielli, A. Crespi, Man-Hong Yung, R. Osellame and A. Peruzzo, Nat.
Comm. {\bf 7}, 11339 (2016).


\bibitem{Kandel2019} Y. P. Kandel, H. Qiao, S. Fallahi, G. C. Gardner, M. J. Manfra, and J. M. Nichol, Nature {\bf 573}, 553 (2019).

\bibitem{Kandel2021} Y. P. Kandel, H. Qiao, and J. M. Nichol, Appl. Phys. Lett. {\bf 119}, 030501 (2021).

\bibitem{Martins2017} F. Martins, F. K. Malinowski, P. D. Nissen, S. Fallahi, G. C. Gardner, M. J. Manfra, C. M. Marcus, and F. Kuemmeth,
Phys. Rev. Lett. {\bf 119}, 227701 (2017).

\bibitem{Baum2022} E. Baum, A. Broman, T. Clarke, N. C. Costa, J. Mucciaccio, A. Yue, Yuxi Zhang, V. Norman, J. Patton, M. Radulaski,
and R. T. Scalettar, Phys. Rev. B {\bf 105}, 195429 (2022).

\bibitem{Kostak2007}V. Kostak, G. M. Nikolopoulos, and I. Jex, Phys. Rev. A {\bf 75}, 042319 (2007).


\bibitem{Zajac2016}D. M. Zajac, T. M. Hazard, X. Mi, E. Nielsen, and J. R. Petta, Phys. Rev. Applied {\bf 6}, 054013 (2016).


\bibitem{Li2018} X. Li, Y. Ma, J. Han, T. Chen, Y. Xu, W. Cai, H. Wang, Y. P. Song, Z-Y. Xue, Z-q. Yin, and L. Sun, Phys. Rev. Applied {\bf 10}, 054009 (2018).

\bibitem{serraGO} P. Serra, A. Ferrón, O. Osenda, Phys. Lett. A {\bf 449}, 128362 (2022).

\bibitem{Dirichlet} G. H. Hardy, E. M. Wright, D. R. Heath-Brown, and J. H. Silverman{\em An Introduction to the Theory
of Numbers}, 6th edn (Oxford University Press) Theorem 200 (2008).


\end{thebibliography}
\end{document}